\documentclass[%
 reprint,
 amsmath,stmaryrd,amssymb,textcomp
 aps,
pra,
]{revtex4-1}
\usepackage{graphicx}
\usepackage{dcolumn}
\usepackage{bm}
\usepackage[draft]{hyperref}
\usepackage{amsmath}
\usepackage{textcomp}
\usepackage{float}
\usepackage{subfig}
\usepackage{wrapfig}
\usepackage{fancyhdr}
\usepackage[justification=centerlast]{caption}
\begin{document}
\title{Dual-mode optical fiber-based tweezers for robust trapping and manipulation of absorbing particles in air}

\author{Souvik Sil\textsuperscript{\dag{}}}
\author{Tushar Kanti Saha\textsuperscript{\dag{}} }
\author{Avinash Kumar}
\author{Sudipta K. Bera}
\author{Ayan Banerjee\textsuperscript{{*}}}

\affiliation{Department of Physical Sciences, Indian Institute of Science Education
and Research, Kolkata, India - 741246}
\affiliation{\dag These authors have equal contributions in the work.}
\affiliation{*Corresponding author: ayan@iiserkol.ac.in}

\begin{abstract}
We develop an optical tweezers system using a single dual-mode optical fiber where mesoscopic absorbing particles can be trapped in three dimensions and manipulated employing photophoretic forces. We generate a superposition of fundamental and first order Hermite-Gaussian beam modes by the simple innovation of coupling a laser into a commercial optical fiber designed to be single mode for a wavelength higher than that of the laser. We achieve robust trapping of the absorbing particles for hours using both the pure fundamental and superposition mode beams and attain large manipulation velocities of $\sim$5 mm/s in the axial direction and $\sim$0.75 mm/s in the radial direction. We then demonstrate that the superposition mode is more effective in trapping and manipulation compared to the fundamental mode by around 80\%, which may be increased several times by the use of a pure first order Hermite-Gaussian mode. The work has promising implications for trapping and spectroscopy of aerosols in air using simple optical fiber-based traps.
\end{abstract}



\maketitle

Optical fiber traps have several advantages over conventional microscope-based traps - the most important being that of large working distance, portability, and ease of optical alignment. However, there exists an issue with employing fiber traps using the optical gradient force - namely, the destabilizing effects of scattering which renders 3-D trapping of particles rather difficult. To overcome  this, a variety of methods have been proposed including dual fiber traps to cancel scattering \cite{Liu09}, specially fabricated fibers such as lensed \cite{cabrini06, ashleigh12}, tapered \cite{zhihai06, jianbin15}, or micro-machined \cite{natphoton07, mohanty08} optical fibers, etc. All these techniques detract from the easy deployment of fibers for optical trapping, so that there remains a limitation in their widespread usage. Moreover, almost all fiber traps developed presently are designed for trapping in liquids since the associated high viscosity results in rather slow rates of particle diffusion. This is not the case in air where the intrinsic low viscosity leads to much faster diffusion, thus requiring considerably steeper optical potentials to achieve stable traps. Optical trapping with fibers by employing gradient forces thus becomes all the more difficult. Recently, photophoretic forces \cite{jovanovic2009photophoresis} have provided an alternate route for trapping absorbing mesoscopic particles in air, since these forces are much stronger than dipole forces and thus considerably relax the conditions necessary for achieving strong traps. 

In  earlier work, we have demonstrated \cite{bera16} stable trapping using photophoretic forces generated by a fundamental Gaussian beam emanating from a laser, so that one could envisage the development of similar traps using single mode optical fibers, which would obviously lead to significant advantages in simplicity of design as well as portability.  In this paper, we describe our efforts towards developing such single mode fiber traps for trapping particles in three dimensions and also improving trap robustness by a simple innovation. Note that in our experimental design, absorbing particles are trapped when they fall against gravity in the sample chamber. The longitudinal photophoretic force acts in the direction of the light beam, so that particles get trapped when this force balances gravity. Thus, particles are often not trapped at the beam center but at that intensity region where gravity is balanced. The radial trapping, however, happens due to a restoring force generated by the helical motion of particles caused by the transverse photophoretic body force, which applies a torque on particles due to its interaction with gravity \cite{Rohatschek1995}. This results in the particle trajectories to be radially shifted off-axis with respect to the trapping beam center, so that they are finally trapped off-axis in the radial direction  with the trap stiffness linearly proportional to intensity \cite{bera16}. In order to maximize the trapping force in the radial direction, it would thus be useful if a beam having high off-axis intensity - such as that found in higher order Hermite-Gaussian transverse mode profiles - could be generated. This is what we achieve in this work. We develop a single fiber based three dimensional optical trap using photophoretic forces, and also demonstrate that the trapping efficiency may be improved by using a first order Hermite-Gaussian beam. We generate such a beam from the fiber itself by coupling light of a wavelength at which the fiber can sustain the first two transverse propagating modes inside it. Thus, we are able to generate a superposition of fundamental and first order Hermite-Gaussian modes by changing the coupling angle of light into the fiber. The trapping efficiency is improved by around 1.8 times which is manifested both in the trapping $Q$ parameter, and the threshold laser power required for trapping. 
\begin{figure}\centering
\includegraphics[scale=0.3]{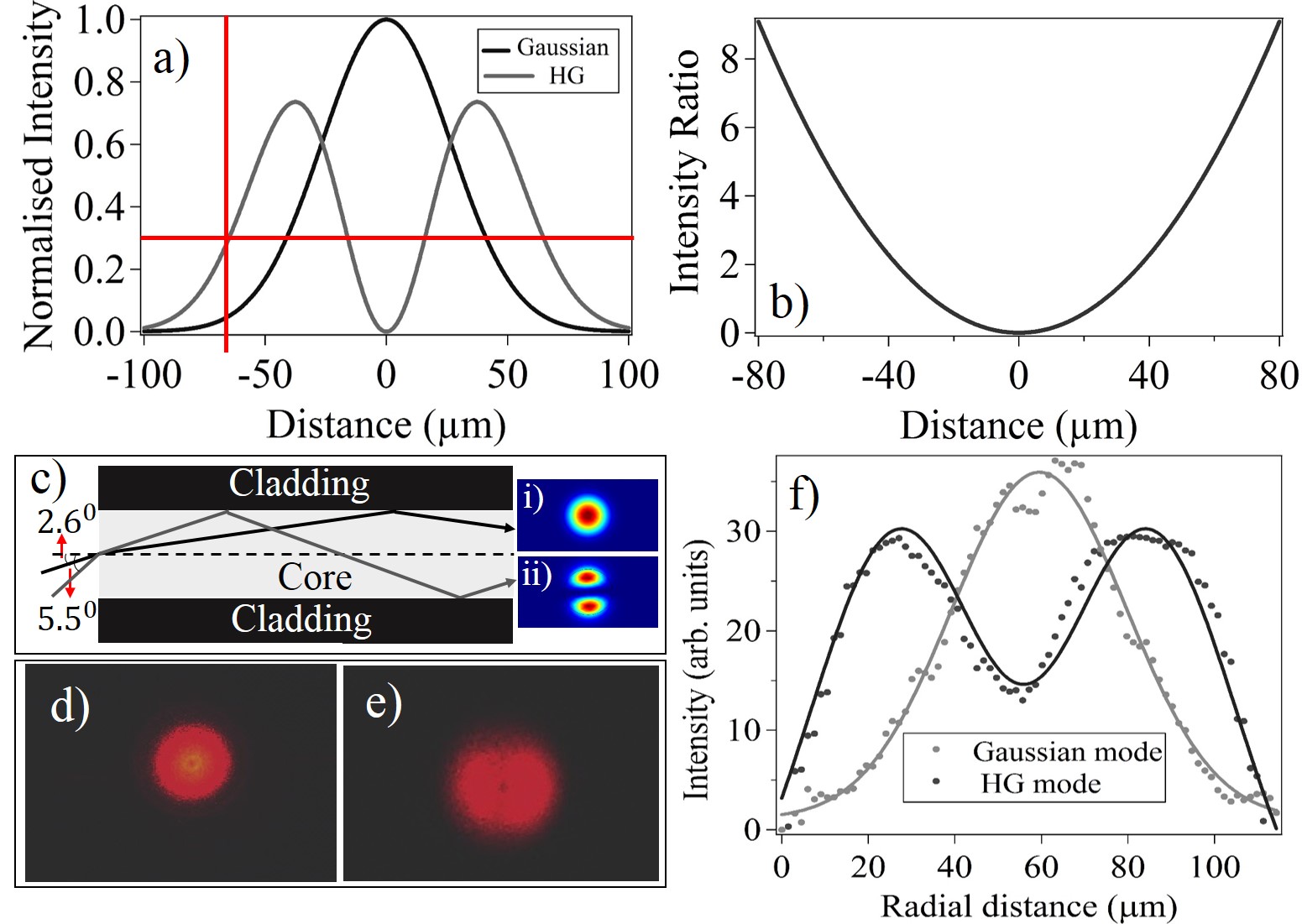}\caption{\label{fig1}a). Theoretical plots of fundamental Gaussian (black) and first excited (grey) modes with beam waist of 53 $\mu$m which is the waist size at an axial distance of 1.5 mm away from the focus. b) Intensity ratio of the Gaussian and first excited modes in a) as a function of radial distance. c) Schematic of beam coupling design to generate fundamental (black, inset (i)) and first excited Hermite-Gaussian (grey, inset (ii)) modes. A larger input angle is required for exciting the  first order mode. d) Image of the fundamental mode obtained experimentally. e) Image of the superposition of fundamental + first excited modes that we use in our  experiments. f) Scatter plots of modes obtained in e) (black) and f) (grey) at 1.5 mm away from the focus with corresponding line fits.}
\end{figure}

We now present a quantitative description of our approach. The solution of the wave equation inside a wave guide (optical fiber) is typically in the form of $TE_{mn}$ ($m,~n$ being positive integers) modes, of which ${TE_{00}}$ and ${TE_{01}}$ (or $TE_{10}, TE_{11}$) are the fundamental and first excited mode (which we refer to as Gaussian and HG henceforth), respectively. Fig. \ref{fig1}a) shows a theoretical plot of the Gaussian (black) and HG (grey) modes for a beam waist size of 53 $\mu$m, which is that corresponding to the mean axial position where particles are trapped in experiments - about 1.5 mm away from the beam focus \cite{bera16}). As is obvious, the transverse extent and off-axis intensity of the first excited mode are both larger compared to the fundamental mode. We plot the intensity ratio of the two beams as a function of radial distance in Fig.~\ref{fig1}b). It is clear that as one moves farther off-axis, the intensity ratio increases, which implies a correspondingly stronger relative trapping force. However, since the overall intensity reduces off-center as well, it is only possible to trap up to a normalized intensity of 0.3 (indicated by red lines in Fig.~\ref{fig1}a)) which corresponds to a radial distance of around 70 $\mu$m, where the HG-beam based trap is close to 6 times stronger in intensity than the Gaussian beam trap. Fig.~\ref{fig1}c) shows the schematic of obtaining the Gaussian and HG modes through the fiber, which we proceed to understand in a more quantitative fashion. On solving Maxwell's wave equations for the electric field propagating in a cylindrical wave guide (the closest approximation of a fiber), one obtains the following equations for the radial component of the field \cite{Ghatak}:
\begin{eqnarray}
E(r,\phi)& =
\dfrac{A}{J_l(U)} J_l\left (\dfrac{Ur}{a}\right )\begin{pmatrix}
\cos(l\phi)\\ 
\sin(l\phi)
\end{pmatrix}, r < a \nonumber \\
&=\dfrac{A}{K_l(W)} K_l\left (\dfrac{Wr}{a}\right )\begin{pmatrix}
\cos(l\phi)\\
\sin(l\phi)
\end{pmatrix}, r > a,
\end{eqnarray}
where, $a$ is the fiber core radius, $J_l$ and $K_l$ are the Bessel and Hankel functions, respectively, and $l$ is an integer. Using the continuity of $\frac{\partial E}{\partial r}$ at the core-cladding interface, one obtains the following conditions:
\begin{eqnarray}
V\sqrt{1-b}\frac{J_{l-1}[V\sqrt{1-b}]}{J_{l}[V\sqrt{1-b}]} = & -V\sqrt{b}\frac{K_{l-1}[V\sqrt{b}]}{K_{l}[V\sqrt{b}]} ,& (l\geq 1) \label{fibereq1}\\
V\sqrt{1-b}\frac{J_{1}[V\sqrt{1-b}]}{J_{0}[V\sqrt{1-b}]} =& V\sqrt{b}\frac{K_{1}[V\sqrt{b}]}{K_{0}[V\sqrt{b}]} , & (l=0)
\label{fibereqn}
\end{eqnarray}
where, $\lambda$ is the wavelength of the laser beam, $n_1$ and $n_2$ are the refractive indices of the fiber core and cladding, respectively, $\left[ V = \frac{2\pi}{\lambda}a\sqrt{n_1^2-n_2^2}\right] $ is the wave-guide parameter,   $\beta$ is the propagation constant of a particular mode, $\left[b = \frac{\beta^2 - k_0^2n_2^2}{k_0^2(n_1^2 - n_2^2)}, ~ (0<b<1) \right]$ is the normalized propagation constant, and $k_0$ is the wave number in vacuum. Now, for a given fiber, one determines the value $V$, and thereby the number of allowed modes from the conditions given by Eq.~\ref{fibereqn}. For our fiber (SM980-5.8-125- Thorlabs) and trapping laser wavelength of 671 nm,$V=3.58$, while the value of $b$  is determined by finding the intersection points of the rhs and lhs of Eqs.~\ref{fibereqn} and ~\ref{fibereq1} plotted against $b$, from which one also determines the values of the mode propagation constant $\beta$ for both modes. Thus, we find that while Eq.~\ref{fibereqn} holds, Eqn.~\ref{fibereq1} is satisfied only for the $l=1$ mode, so that we have the fundamental and first excited Hermite-Gaussian (HG) modes propagating in the fiber.  In addition, the optimal incident angles for both modes are determined from $\theta = \cos^{-1}{(\frac{\beta}{k_0 n_1})}$, as  shown in  Fig.~\ref{fig1}c) - 2.6$^o$ and 5.5 $^o$ for Gaussian and HG modes, with the respective simulated intensity profiles shown in insets (i) and (ii). Thus, it should be possible to easily switch between the two modes by changing the coupling angle into the fiber which we achieve using simple mirrors as shown later. For our experiments, however, we use a superposition of the fundamental and first excited modes since the intensity of the pure first excited mode is too small to trap particles. This is because the ratio of power coupled into the respective modes depends on $log_{10}\left (\frac{NA_{G}}{NA_{HG}}\right )$ \cite{Ghatak}, where $NA_G$ is the numerical aperture (sine of the coupling angle shown in Fig.~\ref{fig1}c)) for the fundamental mode, while $NA_{HG}$ is that for the first excited mode. For our case, this comes out to be around 0.32.  We demonstrate experimentally obtained images of the two types of beams in Figs.~\ref{fig1}d) and e), where the former is the fundamental mode while the latter figure is a representative superposition mode that we typically use in our experiments. We then show scatter plots of the beam profiles in Fig.~\ref{fig1}f) which we fit using standard Gaussian and a superposition of Gaussian + HG  functions with appropriate weight factors which we describe later.
\begin{figure}\centering
\includegraphics[scale=0.27]{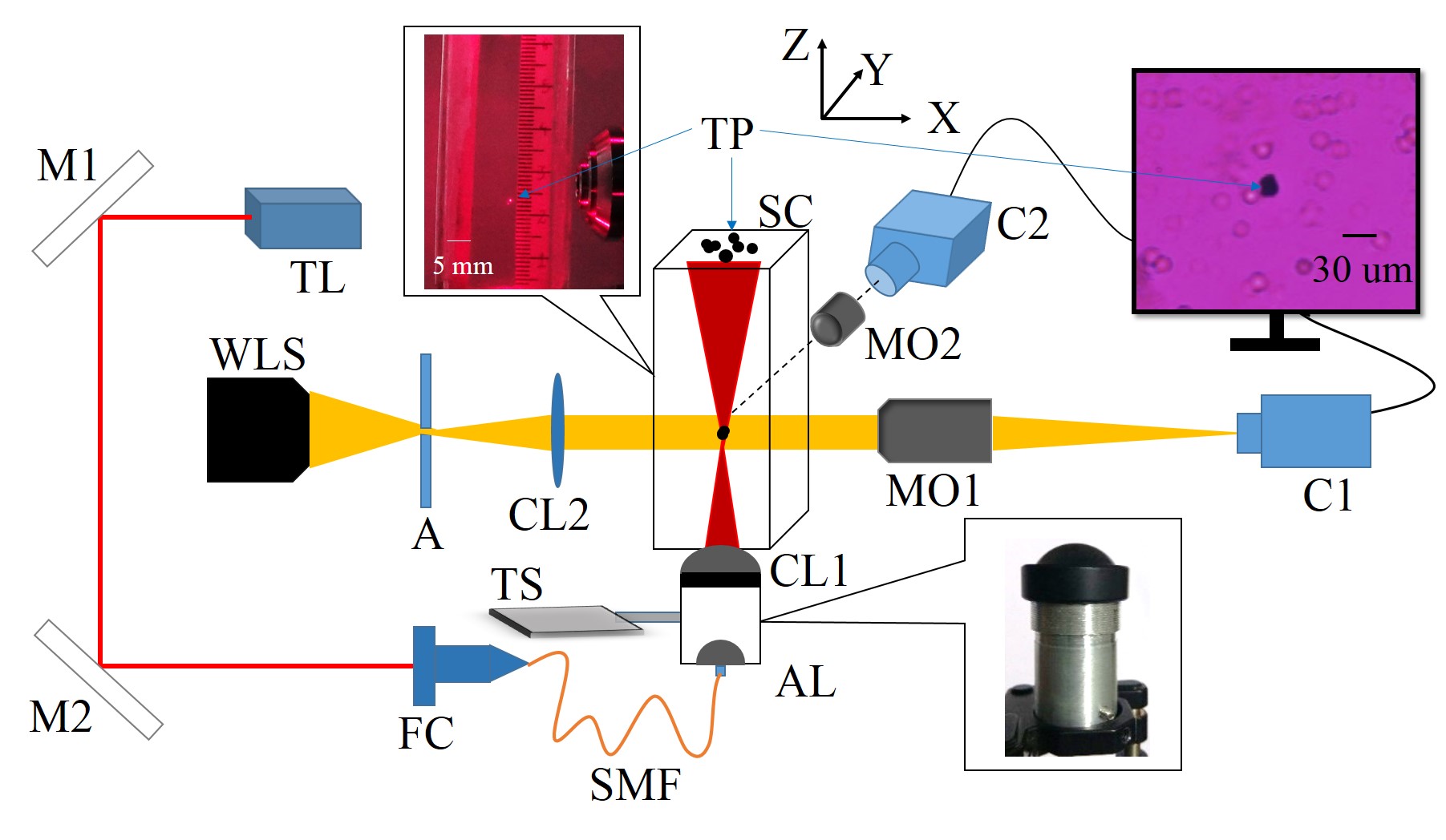}\caption{\label{fig2} Schematic of the experiment. M1 and M2: plane mirrors; TL: trapping laser (671 nm); WLS: White light source; A: Aperture; CL1, CL2: plano-convex lenses; AL: Aspheric lens; FC: Fiber coupler; SMF: Single mode fiber for 980 nm;  MO1 and MO2: 10X objective lenses; SC: Sample chamber; TS: Translation stage; TP: Toner particles; C1 and C2: Cameras. The left top inset shows a zoomed-in image of the sample chamber with a trapped particle and a measuring scale at the background for determining the axial position, while the bottom right inset depicts the actual image of the home-built mount for collimating the SMF output using AL and focusing into SC using CL1.}
\end{figure}

\begin{figure}[!h!t]\centering
\includegraphics[scale=0.32]{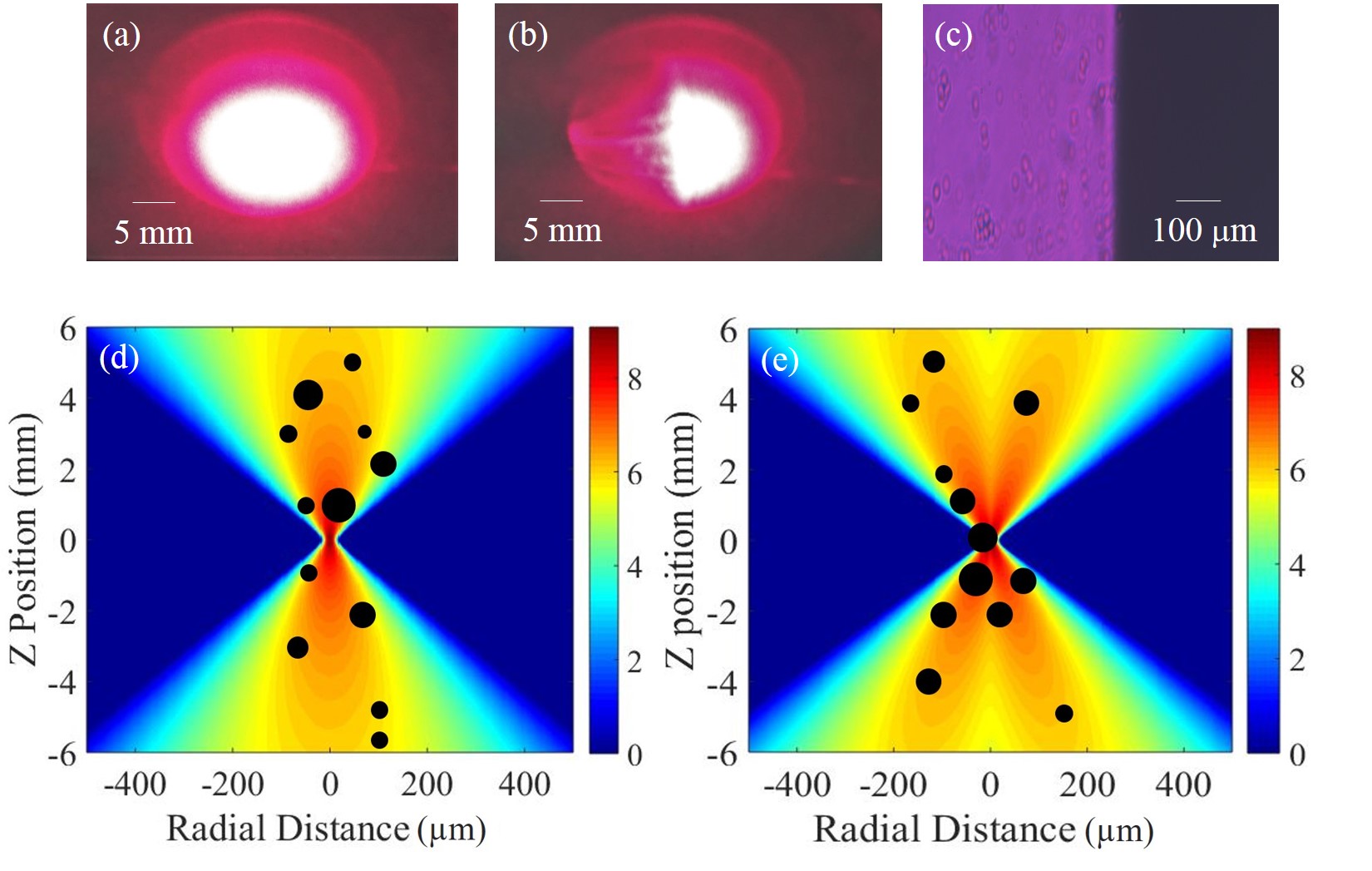}\caption{\label{fig3}a) Beam spot imaged by C3 at the top surface of SC. b) Beam spot after half of it is cut by knife edge inserted into the SC. c) Image of knife edge by C1 (C2). The center line gives the $x=0$ ($y=0)$ line for determining radial coordinates of the beam. d) Trapping locations of particles in Gaussian and e) Gaussian + HG beams after the radial and axial positions are determined using the methods described in the text. Particle are not drawn to scale, however their size ratio has been maintained.} 
\end{figure}

The experimental schematic is shown in Fig.~ \ref{fig2}. For the experiment, we use printer toner particles (TP), a single trapped specimen of which is shown in the inset of the computer in Fig.~\ref{fig2}. The particles are roughly spherical and their average diameter is close to 20 $\mu$m (see \ref{table1}). The trapping laser is a diode laser at 671 nm with maximum power 300 mW which we couple into a single mode fiber SMF using coupling mirrors M1 and M2 and fiber coupler FC. The fiber output is collimated and focused into the glass sample chamber SC (dimensions 75x25x25 mm) via a home-built mount which comprises of an aspheric lens AL for collimation, and a 25 mm plano-convex lens CL1 for focusing (bottom right inset of Fig.~\ref{fig2}). This mount is then attached to a motorized stage TS which, when translated, moves the trapping beam inside SC. The trapped particles are imaged using a white light source WLS which is collimated by lens CL2 and imaged on camera C1 in the $x$-direction and C2 in the $y$ direction using 10x collection objectives MO1 and MO2. For determination of the axial location of a trapped particle we take simultaneous images using another camera C3 (not shown in Fig.~\ref{fig2}) - a representative zoomed in image of the sample chamber with a trapped particle taken using C3 is shown in the top left inset of Fig.~\ref{fig2}. The scale affixed to the sample chamber, and subsequent pixel-by-pixel analysis of the images taken by C3, enables us to measure the axial trapping position to 0.1 mm accuracy. For determining the radial position of the particle, we calibrate the field of views of C1 and C2 by inserting a knife-edge inside SC, and taking simultaneous images by C3 and C1 (C2) for $x$ ($y$) calibration. We first move the knife-edge till it cuts half of the beam (as determined by the images taken using C3 shown in Figs.~\ref{fig3}(a) and (b)), and then obtain the image of the knife-edge at this position using C1 (C2) to determine the $x=0$ ($y=0$) line in the beam. A careful pixel to pixel comparison between different images where the cameras are always held fixed then allows a reasonable localization of the particle with respect to the trapping beam both radially and axially.
\begin{table*}
\caption{\label{table1}Comparison of trap parameters for Gaussian and  Gaussian + HG trapping beams.}
\label{coeff_table} %
\centering
\begin{tabular}{|c|c|c|c|}
\hline
Trap parameters & Gaussian (1) & Gaussian + HG (2) & Ratio (2)/(1)  \\  \hline
Laser Intensity $(MW/m^2)$ & $2.08(42)$ & $3.94(79)$ & 1.89(80)   \\ \hline
Average particle  diameter ($\mu$m)& 16.46(1.69) & 17.42(2.3) &1.06(25) \\ \hline
$Q$ parameter&  0.0059(4) & 0.0108(7) & 1.83(24)   \\  \hline  
Threshold trapping power (mW) & 52.99(4.62) & 33.22(4.01) &1.60(34)  \\ \hline
\end{tabular}
\end{table*}

We trap particles using both the pure Gaussian and the Gaussian + HG superposition mode, and study the trap characteristics for both beam modes. Typically, we obtain different coupling efficiencies for different superposition modes, with the highest power we can couple being in the mode with a very small HG component, where we achieve around 55\% coupling. The maximum power in the pure HG mode is only around 25 mW, which is in accordance with the value predicted by the theory. The trade-off we need to perform is naturally between the transverse extent of the mode and the optical power coupled into it, with the understanding that the maximum transverse extent occurs for the pure HG mode, for which the power is lowest. It is therefore important to optimize the extent of superposition of the two modes, and we perform this by measuring the trapping efficiency at different beam profiles. For best results, the mode profile is that shown in Fig.~\ref{fig1}f), with the superposition mode being fit (dark grey solid line) to a function of the form $y = a_0 \exp\left [- \left (\frac{x-x_0}{2\omega}\right )^2 \right ] + a_1\left (\frac{x-x_0}{\omega}\right )^2 \exp\left [- \left (\frac{x-x_0}{2\omega}\right )^2\right ]$, where $a_0$ and $a_1$ are the weight factors for the Gaussian and HG modes, respectively, $\omega$ is the beam waist, and $x_0$ is the beam center. In this fit, we obtain a ratio of 1:4 for $a_0:a_1$. We now show our experimental results with 12 particles trapped using each beam. The maximum power we obtain in the superposition mode is 70 mW, so for a fair comparison, we keep the same power for the pure Gaussian beam as well. The beam waist at the focus of the 25 mm lens is measured to be around 6.1 $\mu$m (not shown), and around 50$\mu$m at an axial distance 1.5 mm away for both beams as is shown in the line fits in Fig.~\ref{fig1}f). This value is in close agreement with the calculated beam waist of 6.6 $\mu$m and 53 $\mu$m, respectively, considering a measured input beam diameter of around 0.8 mm at the lens CL1. Thus, the input parameters for both traps are set to be the same. Since we have a significant spread in the sizes of the toner particles, they are trapped at different locations in the beam since the equilibrium between the longitudinal photophoretic force and gravity is determined by the particle mass (and hence the size). The location of the trapped particles with respect to the beam propagating in the sample chamber is shown in Figs.~\ref{fig3} (c) and (d). While the particles are not drawn to scale, the relative size ratios have been maintained. It is clear that for both Gaussian (Fig.~\ref{fig3}a)) and HG (Fig.~\ref{fig3}b), the following traits hold: a) particles are always trapped off-center, b) larger particles are trapped comparatively closer to the beam axis since they need to sample the larger intensity regions of the beam so that the resultant photophoretic force may balance gravity, c) the axial extent is similar for both types of beams - around $\pm 5$ mm from the beam center. 

In our experiments, we achieve stable trapping for even up to an hour if perturbations due to air flows are minimized by sealing the sample chamber. It is important to point out that the axial trapping efficiency is similar for both pure and superposition mode beams. We are able to impart axial velocities of 5 mm/s to the trapped particles without losing them from the trap. This is achieved by translating the trap itself by vertically moving TS as mentioned earlier. The comparison of the trap parameters for radial trapping using the pure and superposition modes is shown in Table 1, with the numbers in parenthesis denoting $1\sigma$ errors in the mean. We first quantify the light intensity at the trapped particle locations which is determined from the positions of the particles  from Fig.~\ref{fig3}(c) and (d) and the knowledge of the beam waist and the laser power coupled into the trap. Thus, the average intensity sampled by particles in the Gaussian + HG beam is greater than that by the Gaussian beam by a factor of 1.89(80). The next parameter in Table 1 is the average particle size where we observe that the difference in particle sizes is not statistically significant ($p$-value = 0.76 in a 2 parameter t-test). We then determine the trapping force by the drag force method, where we impart an acceleration of 0.1 $mm/s^2$ to TS, so that the trap is translated radially. We measure both the time taken and distance traversed by the particle before it leaves the trap. Both time and distance yield values of particle velocity at the point of escape from Newton's laws of motion, and they agree to within 5\% of each other. Thus, the velocity $v$ attained by particles in the Gaussian trap is determined to 0.43(3) $mm/s$, while that in the Gaussian + HG beam is 0.75(5) $mm/s$, so that the trapping force $F$ is calculated to be around 1.39(9) pN for the Gaussian beam and 2.49(18) pN for the Gaussian + HG beam using $F = 6\pi \eta r v$ where $r$ is the radius of the particles (from Table 1), and $\eta$ = viscosity of air, which we determine at  50 degC from Ref.~\cite{engineering}. This value of temperature is found by observing that the toner particles melt at a laser power of 120 mW, for which the temperature of the particles is around 85 degC (measured using a thermocouple).  Thus, we can determine the $Q$-parameter of the traps for Gaussian and HG beams from $Q = Fc/nP$, where $c$ is the velocity of light, $P$ is the laser power, and $n$ is the refractive index of air which we can consider to be 1 in our case. This comes out to be 0.0059(4) for the Gaussian beam, and 0.0108(7) for the Gaussian + HG beam, which shows that the trapping efficiency is around 1.83(24) times higher for the latter. Finally, we show that the threshold power for trapping is around 52.99(4.62) mW for the Gaussian beam and 33.22(4.01) mW for the Gaussian + HG beam, so that the latter can trap at around 60\% less power than the former. Note that by threshold power, we signify the power at which particles leave the trap after being initially trapped at higher powers. This indicates the Gaussian + HG beam trap is about 60\% more stable than the Gaussian beam trap for toner particles. 

In conclusion, we employ photophoretic forces to develop a three dimensional trap for absorbing particles using a single optical fiber. The fiber is dual mode at the laser wavelength of 671 nm, and we can controllably excite both a pure Gaussian and a superposition of a Gaussian and first order Hermite Gaussian modes in the fiber to trap single printer toner particles. The axial trapping is robust for both types of beams, and we achieve particle translation velocities of 5 $mm/s$ axially. The superposition beam appears to be more efficacious in radial trapping with around 80\% higher trapping efficiency, a number which is corroborated from measurements of the maximum radial translation velocities of the trapped particles (0.43(3) $mm/s$ for the Gaussian beam and 0.75(5) $mm/s$ for the Gaussian + HG beam), intensities of the beams where the particles are trapped in three-dimensions, and the threshold trapping power. This work shows that photophoretic forces can lead to new paradigms in developing simple but robust, portable single fiber-based optical traps enabling spectroscopy of aerosols/bioaerosols, and other diverse applications. 

This work was supported by the Indian Institute of Science Education
and Research, Kolkata, an autonomous research and teaching institute
funded by the Ministry of Human Resource Development, Govt. of India.

\end{document}